\documentclass[12pt]{article}
\usepackage{mathtext}
\usepackage{graphicx}
\usepackage{amsfonts}
\usepackage{amssymb}
\usepackage{amsmath}
\numberwithin{equation}{section}

\newcommand{\nn}{\medskip}

\renewcommand{\d}{\delta}

\newcommand{\ar}{\longrightarrow}
\newcommand{\w}{\omega}
\newcommand{\s}{\sigma}
\newcommand{\la}{\lambda}

\begin{document}
\title{Space of dark states in Tavis-Cummings model}
\author{Yuri I.Ozhigov\thanks{ozhigov@cs.msu.su}\\
Moscow State University of Lomonosov, VMK, \\
Institute of physics and technology RAS}
\maketitle

\begin{abstract}
The dark states of a group of two-level atoms in the Tavis-Cummings resonator with zero detuning are considered. In these states, atoms can not emit photons, although they have non-zero energy. They are stable and can serve as a controlled energy reservoir from which photons can be extracted by differentiated effects on atoms, for example, their spatial separation. Dark states are the simplest example of a subspace free of decoherence in the form of a photon flight, and therefore are of interest for quantum computing. It is proved that a) the dimension of the subspace of dark states of atoms is the Catalan numbers, b) in the RWA approximation, any dark state is a linear combination of tensor products of singlet-type states and the ground states of individual atoms. For the exact model, in the case of the same force of interaction of atoms with the field, the same decomposition is true, and only singlets participate in the products and the dark states can neither emit a photon nor absorb it. The proof is based on the method of quantization of the amplitude of states of atomic ensembles, in which the roles of individual atoms are interchangeable. In such an ensemble there is a possibility of micro-causality: the trajectory of each quantum of amplitude can be uniquely assigned.
\end{abstract}
\section{Introduction. Background}

Interaction between light and matter described by quantum electrodynamics (QED) is the most fundamental force, and at the same time it represents the simplest illustration of the power of quantum theory (see \cite{Fe1},\cite{Fe2}) in its single-particle form, described by the Feynman diagrams. From a logical point of view, fully justified is quantum electrodynamics of a single charge, which can be renoralized by the theorem of Bogolubov and Parasuk (see \cite{BoP} and also \cite{An}). 

For the many body quantum electrodynamics the corresctness rests not on the possibility to renormalize it but rather on the adequacy of the transition to tensor products of spaces of states that by default is considered an absolutely legal mathematical technique for systems of many bodies. 
This method never failed in cases where we could calculate the amplitude of the transition to the end, and gave predictions surprising on the accuracy. However, extrapolation of this technique to systems of many non-identical charges can not give any verifiable result due to the exponential growth of computational complexity with increasing number of charges. This led to the fundamental idea of ​​a quantum computer (\cite{Fe3}), as a necessary tool for modeling complex multi-charge systems. A quantum computer with computational capabilities goes beyond the scope of the computational apparatus of physics accessible to us (fast quantum computation - see \cite {Gr}), and therefore its very idea needs a particularly careful experimental verification and necessary refinements.

The results of numerous experiments conducted since the early 1980s showed that it is hardly possible to build a quantum computer straightforwardly according to the original Feynman scheme (\cite{Fe3}) because of the decoherence phenomenon associated with the inability to isolate the quantum system from the medium (a review of approaches to open quantum systems, see the book \cite{BP}). Therefore, the problem of finding quantum states that would be isolated from the medium by its very form and would have sufficient flexibility to map all quantum states in general (a known attempt in this direction is a topological quantum computer, see \cite{FCL}) has come to the forefront. 

In this paper we study the simplest states of ensembles of two level atoms: dark states. It is proved that such states are exclusively superposition of tensor products of EPR singlets, e.g. states of the form $|01\rangle-|10\rangle$. This means that optical darkness for two-level systems is closely related to the spin description: singlet states have zero total spin. Such a transparent connection exists only for two-level systems, that is, for spin 1/2.

Another aspect of the problem of quantum computers is overcoming the computational difficulties that inevitably arise when applying QED to the modeling of quantum computing. Quantum computation itself can be performed on the states of charged particles (spatial positions or spins), but the main source of decoherence is the interaction of charges with the field. Therefore, the simulation of a quantum computer must take place within the framework of QED, which is much more complicated than ordinary quantum mechanics, in which the field is manifested only in the form of a scalar potential.

Of particular importance are finite-dimensional models of QED, in which it is possible to reduce the complex states of the electromagnetic field to several qubits, meaning the presence or absence of a photon of a certain mode in a limited space-time region. The main of these models was proposed by Jaynes and Cummings for a two-level atom located in an optical Fabry-Perot resonator \cite{JC}), and then was generalized to ensembles of such atoms (the Tavis-Cummings or Dick-
see \cite{D}) and on several cavities connected by an optical fiber (the Jaynes-Cummings-Hubbard model \cite{JCH}). Within these models and their multiple options, it is possible to describe accurately the effects important for applications, for example, DAT (dephasing assisted transport - \cite{HP},\cite{DAT}). On the basis of finite-dimensional models of QED it is possible to obtain nonlinear optical effects, which in principle opens door to construction of elementary gates for quantum computations (see \cite{A}).The JCH model serves as an important generalization of the so-called continuous quantum walks (\cite {Am}) and can be used for their practical implementation.

The states of atoms with nonzero energy, in which they do not emit a photon are called dark states. Such states are not subject to decoherence because, even if they have a high energy of atomic excitations, they can stay in this state theoretically indefinitely for a long time without emitting photons. For two-level atoms, such states can be obtained in an optical cavity, for example, using the Stark-Zeeman effect
 (\cite{Oz}). 

It is possible to extract energy in the form of photons from an atomic system in a dark state by spatial separation of atoms, dephasing noise or other differentiated impacts to atoms. In this case, the resonator is needed only to obtain a dark state, the atomic system can be then removed from the cavity, while retaining the property of darkness, provided that we keept atoms together (for example, using optical tweezers).

Dark states have numerous uses. In particular, their role in the organization of inter-atomic interaction was considered in the work \cite{AL}, for the control of solid-state spins - in work\cite{HS}, 
for the control of macroscopic quantum systems - in work
 \cite{LG}, 
one of the effects of the dark state in the light-harvesting complex can be found in the work \cite{FH}.  
Some methods for obtaining dark states in quantum dots can be read in papers  \cite{PB}, and also in \cite{T}. 
The destruction of dark states by a magnetic field or modulated laser polarization was considered in
 \cite{B}. In the works \cite{CHI},\cite{T},\cite{CEM} singlet states are also considered as states with zero total spin forming the core of the decreasing operator, however, there is no detailed analysis of the structure of the subspace formed by them in these articles.
 
The purpose of this paper is an explicit description of the of dark states. It follows from their definition that they form a subspace, which we will call dark subspace. We will be interested in the structure of this subspace and its dimension. The structure of dark states in the systems of kudits ($ d $ -two systems) is most thoroughly studied in the work \cite {K}. In particular, for two-level systems in the work
\cite{K} it is proved that the dark states are precisely the stationary points of the tensor product of the groups $ SU (2) $. These stationary points are called in this work ''singlet states'', since two-atom singlets of the EPR-pair type $|01\rangle-|10\rangle$ are invariant for this group. 

We shall prove that the dark states can be represented as a linear combination of products of simple singlets, that is, tensor products of EPR pairs. This fact justifies the term ''singlet state'', having a chemical origin: singlet states of electron spins are pairing for atoms, that is, they make it possible to form a covalent bond.

We consider Tavis-Cummings model, consisting of the optical cavity - the resonator, and a group of identical two-level atoms inside it. The cavity length $ L = \pi c / \omega_c $ is equal to half the wavelength of a photon with a frequency
$\omega_c$, which differs from the frequency of atomic transition $\omega_a$ by the small detuning $\d=\omega_c-\omega_a,\ |\d |\ll\omega_c$. 
A small detuning value provides a constructive interference of the electric field of the photons inside the cavity and a long retention time of the photons of frequency
 $\omega_c$ inside the cavity. 

In this case, we can write the Hamiltonian of the interaction of atoms and the field inside the cavity in the dipole approximation in the Jaynes-Tavis-Cummings form:
\begin{equation}
H_{TC}=\hbar\omega_ca^+a+\hbar\omega_a\sum\limits_{j=1}^n\s_j^+\s_j+H_i,\ \ H_i=
\sum\limits_{j=1}^ng_j(a^++a)(\s_j^++\s_j),
\label{TC}
\end{equation}
where $^+$ means conjugation, $a^+,a$ are field operators of creation - annihilation of photon, $\sigma^+_j,\sigma_j$ are raising and lowering operators of $j$-th atom, acting on its ground ($|0\rangle_j$) and excited ($|1\rangle_j$) states as $\s_j |0\rangle_j=0,\ \s_j|1\rangle_j=|0\rangle_j$ (here and below, by default, it is assumed that the remaining state components are acted upon by the identity operator
). Here the force of interaction of an individual atom $j$ with the field $g_j=d\ E_j\sqrt{\hbar\omega_a/2\varepsilon_0L}$, $E_j= sin(\pi x_j/L)$ is the distribution of the photon field intensity along the resonator ,  $x_j$ is the coordinate of the atom along the axis of the cavity, $V$ is the effective cavity volume, $d$ is the dipole moment of an atom, $\varepsilon_0$ is the electric constant. We suppose, for simplicity, that the detuning $\w_c-\w_a$ is zero. The frequencies and strength of the interaction are always assumed to be nonzero.

We denote the part of the interaction of the Hamiltonian of the form
$\sum\limits_{j=1}^ng_j(a^+\s_j+a\s^+_j)$ by $H_{RWA}$, and the other part of interaction $\sum\limits_{j=1}^ng_j(a^+\s_j^++a\s_j)$ by $H_{nonRWA}$.

In the case of weak interaction $g_j/\hbar\omega_a\ll 1$ we can leave ony summands $a^+\s_j$ и $a\s_j^+$, conserving the energy, e.g. $H_{RWA}$ and the other two, which do not conserve the energy $H_{nonRWA}$, we can omit (rotating wave approximation RWA). 

A state that can emit a photon is called a bright state. A state, which is not a bright will be thus dark (see, for example, \cite{FB}). A state that can not absorb a photon, we call transparent. A transparent dark state we call invisible. In other words: invisible is a state of atoms in the cavity, which can neither emit nor absorb a photon, e.g. the ensemble in this state does not interact with the field.


A complete state of the system of atoms and the field has the form of a superposition of the basic states $|j_p\rangle|j_a\rangle$: $|\Psi\rangle_{gen}=\sum\limits_{j_p,j_a}\la_{j_p,j_a}|j_p\rangle$, where the natural number $j_p$ denotes the number of photons in the field, and the binary string $j_a$ denotes the state of distinguishable atoms taken in a fixed order, so $0$ and $1$ denote the ground and excited states of the corresponding atom. Elements $j_1, j_2,...,j_n$ of the string $j_a=(j_1, j_2,...,j_n)$, uniquely corresponding to atoms, we call qubits. A complete state of the system $ |\Psi\rangle_{gen}$ belongs to the tensor product ${\cal H}={\cal H}_p\otimes{\cal H}_a$ of the state spaces of the field and states of atoms. In this article, we are only interested in processes with the emission of at most one photon, so the main object will be the atomic states having the form $|\Psi\rangle=\sum\limits_{j=0}^{N-1}\la_j|j\rangle_a$, which by default we call states, and the index $a$ we omit.

If we assume RWA approximation, an example of a dark two atomic state can be: $|d_1\rangle = |00\rangle$, an example of a transparent - $|t_1\rangle= |11\rangle$.

We introduce the notation $\bar\s=\sum\limits_q\s_q$.
From the form of the interaction of matter and light it follows that the operator of emission of a photon in the RWA approximation is the action of the operator
 $a^+\bar\s$, and for the exact model - of the operator $a^+(\bar\s+\bar\s^+)$. 
Similarly, the photon absorption operator for the RWA approximation is
 $a\bar\s^+$, and for the exact model it coincides, to within an inversion of the field component, with the photon emission operator: $a(\bar\s+\bar\s^+)$. Therefore, the subspaces of dark and transparent states in the RWA approximation are the kernels of operators  $\bar\s$ and $\bar\s^+$ correspondingly, and the invisible is the intersection of these sets. In the exact model the dark, transparent and invisible states are the same - the kernel of the operator $\bar\s+\bar\s^+$.

So, the properties of darkness and transparency, taken separately from each other, depend on the applicability of RWA approximation to the considered model. The states $ | d_1 \rangle $ and $ | t_1 \rangle $ are dark and transparent only if it is applicable. If we refuse from the RWA approximation, these states will lose these properties. For example, the state $ | d_1 \rangle $ becomes bright if the Hamiltonian has the form \eqref{TC}, since $|0\rangle_p|d_1\rangle$ can go to a state with one photon of the form $\frac{1}{\sqrt{2}}|1\rangle_p(|01\rangle+|10\rangle )$.

Throughout, we will identify the base state $ | j \rangle $ with the string of the binary expansion of the natural number $ j $.

Let us consider an example of two-qubit states in the RWA approximation. First, let the interaction force of both atoms with the field be the same: $ g_1 = g_2 $. We choose as the new basis the triplet and singlet states of the form $ | t_0 \rangle = | 00 \rangle, | t_1 \rangle = | 11 \rangle, | t \rangle = \frac {1} {\sqrt {2}} (| 10 \rangle + | 01 \rangle) $, $ | s \rangle = \frac {1} {\sqrt {2}} (| 10 \rangle- | 01 \rangle) $. From them the singlet alone is invisible, and the triplet is neither dark nor transparent. Now suppose that $ g_1 \neq g_2 $, for example, atoms occupy different positions in the resonator. Then the state $ g_2 | 10 \rangle-g_1 | 01 \rangle $ (the atoms are numbered from left to right) will be dark, the state $ g_1 | 10 \rangle-g_2 | 01 \rangle $ is transparent, and there will be no invisible states at all.

In the future, we consider the case of atoms with the same interaction energy with the field: $g_i=g,\ i=1,2,...,n$, the detuning $\w_c-\w_a$ between the frequencies of atoms and the cavity is assumed to be zero, and we will consider only RWA approximation (unless explicitly stated otherwise), up to the last paragraph, where we consider the general case.

The weight (Hamming) $ \nu_j $ of the basic state $ | j \rangle $ is the number of units in it.
The ground state of the atoms $ | j \rangle $ is called equilibrium if its weight is half the number of all atoms. Equilibrium states, therefore, are possible only for systems with an even number of atoms. The superposition of equilibrium basis states is called the equilibrium state of atoms. A more general property of atomic states is linearity. The atomic state $ | \Psi \rangle $ is linear if all its basic components have the same weight.

We will show that the invisibility property does not depend on the applicability of the RWA approximation, in particular, all invisible states are equilibrium.

\section{Structure of the dark subspace}


Let $ | j \rangle $ be the base state of the system of $ n $ qubits; we introduce the notation $ N = 2^n $ - this is the dimension of the entire quantum state space of the $ n $ - qubit system. We denote by $ 1 (j) $ the Hamming weight of this state, i.e. number of units in it; then the number of zeros in it is $ 0 (j) = n-1 (j) $. We define a binary relation on the basis states, denoted by $ Emission (j, j ') $, which is true if and only if $ j' $ is obtained from $ j $ by replacing the single unit by zero. In other words, $ j '$ is obtained from $ j $ by the action of the decreasing operator $ J_- $ on one of the atoms in the excited state. In this case $1(j')=1(j)-1$.

The emission of a photon by an atomic system in a state
 $|j\rangle$, has the form 

\begin{equation}
|0\rangle_p|j\rangle\ar |1\rangle_p|j'\rangle,
\label{emission}
\end{equation}
where $Emission(j,j')$.

For a basic state $|j'\rangle$ we call $j'$- family the set of basic states $|j\rangle$, such that $Emission(j,j')$ is true. In the other words,  $j'$- family consists of basic states $|j'\rangle$, for which the transition of the form (\ref{emission}) is the photon emission. $j'$- family we denote by $[j']$ and call the state $|j'\rangle$ its parent. 

Note that two different families can have no more than one common member.

Let us now consider an arbitrary atomic state
$|\Psi\rangle=\sum\limits_j\la_j|j\rangle$. From the definition of emission of a photon it follows that the state
 $|\Psi\rangle$ is dark if and only if the system of equations of the form

\begin{equation}
\sum\limits_{s\in [j']}\la_s=0, 
\label{dark_condition}
\end{equation}
is satisfied for all $ j'=0,1,\ldots, 2^n-1.$
Note that it is sufficient to require that these equalities be satisfied only for
 $j'=0,1,\ldots, 2^n-2$, because the family $[2^n-1]$ is empty: no state can pass to the basic state consisting of only excited atoms when the photon is emitted.

We denote by $B^n_k$ the set of basic $n$- qubit states $j$, such that $1(j)=k$, and by ${\cal H}^n_k$ - the subspace spanned on $B^n_k$. Then for any basic state $j'$ its family completely belongs to  $B^n_{1(j')+1}$. Consequently, every dark state is a superposition of dark states belonging to subspaces
 ${\cal H}^n_k$, $k=0,1,\ldots,n-1$. We denote by $D^n_k$ the subspace ${\cal H}^n_k$, consisting of dark states. Then $D^n_k={\cal H}^n_k\cap Ker(\bar \s)$.

We will always number the qubits from left to right, denoting by the symbol $ * $ the missing qubit, so that, for example, instead of
$|0\rangle_1 |1\rangle_3$ we write $|0*1\rangle$.

The examples of states from $D^n_k$ are the so called $(n,k)$-singlets: the states obtained by the tensor product of $ k $ samples of states of the form
$|0\rangle_p|1\rangle_q-|1\rangle_p|0\rangle_q$, where $1\leq p<q\leq n$ and $n-2k$ states of the form $|0\rangle_q,\ 1\leq q\leq n$. For $n=4,\ k=2$ $(n,k)$-singlets will be, for example, the following states 

\begin{equation}
\begin{array}{ll}
&(4,2)_1=(|0*1*\rangle-|1*0*\rangle )(*|0*1\rangle-|*1*0\rangle )=|0011\rangle-|0110\rangle-|1001\rangle+|1100\rangle ),\\
&(4,2)_2=(|0\rangle |1\rangle- |1\rangle |0\rangle)^{\otimes 2}=|0101\rangle-|0110\rangle-|1001\rangle+|1010\rangle,\\
&(4,2)_3=(|0**1\rangle-|1**0\rangle )(|*01*\rangle-|*10*\rangle)=|0011\rangle-|0101\rangle-|1010\rangle+|1100\rangle ).
\end{array}
\label{(4,2)}
\end{equation}

These states will be linearly dependent, but any two of them are linearly independent and form a basis of $ D^4_2 $, which is easy to verify directly.

We note that for $n=2k$ all $(n,k)$- singlets are invisible without RWA.

{\bf Theorem}

{\it 1. \ $dim (D^n_k)=max\{ C^k_n-C^{k-1}_n,\ 0\}$.

2.\ \ Any state from $D^n_k$ is the linear combination of  $(n,k)$- singlets}
\nn

{\bf Proof}

At first we prove the point 1.

Since a state $|\Psi\rangle=\sum\limits_j\la_j|j\rangle$ is dark if and only if the system of equation (\ref{dark_condition}) is satisfied, the belonging $|\Psi\rangle\in D^n_k$ is equivalent to the satisfaction of the system $S^n_k$ consisting of all equalities of the form (\ref{dark_condition}) for all $j'$,such that $1(j')=k-1$. If $k=n$, then $dim(D^n_k)=0$ and point 1 is satisfied; since it is sufficient to consider the case $k<n$. Then to the different $j'$ will correspond the different equations from  $S^n_k$. Since the system $S^n_k$ has $C_n^k$ variables and $C_n^{k-1}$ equations to prove point 1 it would suffice to show that all equations from $S^n_k$ are independent. 

Any permutation of $\pi$ from the group $S_n$ acts naturally on the set $B= \{ 0,1\}^n$ of all binary strings $j$ of length $n$; the result of such action is denoted by $\pi j$. In particular, the substitution $(a,b)\in S_n$ acts as a transposition of two qubits with the numbers $a$ and $b$ of the given string. We will call such a transposition essential if it affected two qubits with the different values. Then those and only those transpositions that change the string on which they act will be essential.

\bigskip

{\it Lemma 0. For any string $j\in B$ and any $\pi\in S_n$ the string $\pi j$ has the form 
$
(a_s,b_s)(a_{s-1},b_{s-1})...(a_1,b_1)j$, where all numbers $a_1,a_2,...,a_s,b_1,b_2,...,b_s
$ are different and $s$ equals the double Hamming distance between $j$ and $\pi j$.}
\bigskip

{\it Proof.} Let $s$ be minimal of such numbers that for some set of substitutions $(a_q,b_q),\ q=1,2,...,s$ the string $\pi j $ has the form $(a_s,b_s)(a_{s-1},b_{s-1})...(a_1,b_1)j$. We prove that all numbers $a_1,a_2,...,a_s,b_1,b_2,...,b_s$ are different. Indeed, let it be wrong and some qubit is affected twice. Since always $(a,b)=(b,a)$ and the substitutions of the form $(a,b)$ and $(c,d)$ for the different $a,b,c,d$ commute we can change the places of substitutions $(a_q,b_q)$ so that two of them $(a_q,b_q),(a_{q-1},b_{q-1})$, such that $b_{q-1}=a_{q}$ becomes ajacent. Since a$s$ re minimal among the numbers of qubits $a_q, b_q,a_{q-1},b_{q-1}$ are exactly 3 different and we can assume that the numbers of qubits a$a_{q-1},a_q,b_{q}$ re different. The values of these qubits in the binary string $j'=$ $(a_{q-2},b_{q-2})(a_{q-3},b_{q-3})...(a_1,b_1)j$ we denote by $a,b,c$. Thanks to minimality of $s$ we have $a\neq b$ and we can assume that $a=0,b=1$. If $c=0$, the substitution $(a_{q},b_{q})$ is undue. If $c=1$, then $(a_q,b_q),(a_{q-1},b_{q-1})j'=(a_{q-1},b_{q})j'$ and the condition of mnimality is violated again. Hence, all qubits participating in the considered substitutions have the different numbers and their values in each substitution are different as well. It involves that $s$ is double Hamming distance between $j$ and $\pi j$. Lemma 0 is proved. 

We define the natural metrics on the set $B^n_{k-1}$ as follows. The distance $d(j,j')$ between basic states $j,\ j'\in B^n_{k-1}$ is defined as the half of Hamming distance between them that is by Lemma 0 is the minimal number of substitutions (permutations of a pair of qubits) in the transition from  $j$ to $j'$.\footnote{So defined distance - through the number of substitutions are more convenient than Hamming because Hamming distance between elements of $B^n_{k-1}$ are always even.} 

 Sequence of substitutions $j_0\ar j_1\ar\ldots j_r$ we call correct if all passages $j_i\ar j_{i+1}$, $i=0,1,...,r-1$ are essential substitutions and any qubit is affected in it no more than once.

We fix the arbitrary $j_0\in B^n_{k-1}$.

{\it Lemma 1. Let $j_0,j_1,j_2,\ldots ,j_r$ be a sequence of states from $B^n_{k-1}$. If for any $q=0,1,\ldots r-1$ 

\begin{equation}
d(j_q,j_0)=d(j_{q+1},j_0)-1,\ d(j_{q+1},j_q)=1,
\label{eq0}
\end{equation}
then there exists the correct sequence of substitutions of the form $j_0\ar j_1\ar\ldots j_r$, in which substitutions are determined uniquely and vice versa, if there exists such correct sequence then for all  $q=0,1,\ldots r-1$ the equalities \eqref{eq0} are true. }

Induction on $r$. The basis is evident. Step. Let Lemma 1 be true for $r-1$ and prove it for $r$. Let at first equations \eqref{eq0} be satisfied. By the induction hypothesis, there exists a correct sequence $P$ of substitutions
 $j_0\ar\ldots j_{r-1}$, and by $d(j_{q+1},j_q)=1$ the passage $j_{r-1}\ar j_r$ - is a substitution as well. This substitution must change zero and one, because otherwise we would have the contradiction with the condition $d(j_{r-1},j_0)=d(j_{r},j_0)-1$. Then, if this step violates the correctness, there is a qubit that participates twice in transpositions from $j_0\ar\ldots\ar j_r$ and it is affected just at the last step $j_{r-1}\ar j_r$. But then we could reduce this sequence of substitutions, having received a contradiction with condition $d(j_q,j_0)=d(j_{q+1},j_0)-1$. Indeed, without loss of generality we can assume that the sequence $P$ moves units from qubits with numbers $1,2,...,r-1$ to the positions $r,r+1,...,2r-2$ in random order, on which initially standed zeroes, and the last substitution $j_{r-1}\ar j_r$ moves the $2r-2$-th qubit either to the place $r-1$, or to the place $2r-1$. In the first case the sequence $P$ can be reduced to sharter since its result can be reached by the mobement of only  $r-2$ qubits. In the second case we can reduce the sequence $j_0\ar j_1\ar\ldots j_r$, because it factually replaces only $r-1$ units by zeroes, and by Lemma 0 it means that  $d(j_{r-1},j_0)=d(j_{r},j_0)$, which contradicts to the condition.

Let the sequence $j_0\ar\ldots\ar j_r$ be a correct sequence and by the inductive hypothesis the equalities \eqref{eq0} are true for all $q=0,1,\ldots,r-1$. The second equality will be true because $j_{r-1}\ar j_r$- is a substitution. If the equality $d(j_{r-1},j_0)=d(j_r,j_0)-1$ is violated then the passage from $j_0$ to  $j_r$ can be fulfilled in less than $r$ substitutions and Hamming distance between $j_0$ and $j_r$ is less than  $2r$ that contradicts to the correctness of the sequence  $j_0\ar\ldots\ar j_r$, because in it each qubit is affected only once and the Hamming distance between $j_0$ and $j_r$ is then $2r$. Lemma 1 is proved. 


We define the partial order on $B^n_{k-1}$, putting  $j_1<j_2$, if and only if there exists the correct sequence of substitutions of the form
$j_0\ar\ldots\ar j_1\ar\ldots\ar j_2$. Then we can arrange all the states in $B^n_{k-1}$ at the nodes of the graph $D$, in the initial vertex of which is $ j_0 $, and for any vertex $ j '$ all vertices $ j $ lying above $ j' $ connected to $ j '$ by an edge satisfy the equalities $d(j,j_0)=d(j',j_0)+1$ and are obtained from $ j '$ by exactly one substitution. In this case, any monotonically increasing path on this graph will contain vertices in increasing order of $d(j,j_0)$. The existence and uniqueness of such a graph $ D $ follows from Lemma 1. We enumerate tiers of this graph beginning with zero tier, consisting of only $j_0$. 
 
The basic states $ j '\in B^n_{k-1} $, lying in the tier $ p $, will be called the parents of rank $ p $. The rank of such a parent is equal to the total number of qubit numbers that are equal to one in $ j_0 $, and zero to $ j '$, that is, the Hamming distance between these vertices. We will denote the set of these qubit numbers by $ rem (j') $. The rank of the state $ j \in B^n_k $ is the minimal rank of the parent $ j '\in B^n_ {k-1} $ whose family contains $ j $: $ j \in [j'] $. The rank of state $ j \in B^n_k $ is denoted by $ r (j) $.

{\it Lemma 2. Let the parent $ j '\in B^n_{k-1} $ have rank $ p $. Then exactly $ p $ of its family members have rank $ p-1 $, the remaining $ n-k + 1-p $ have rank
 $p$.}

{\it Proof}. We first we note that $0\leq p\leq min\{ k-1,n-k+1\}$. It follows from the definition of the rank of the elements $ B^n_k $ that the members of the family $ [j '] $ having rank $ p-1 $ are exactly the basic states $ j $ obtained from $ j' $ by replacing zero by a unit in some qubit from $ rem \ (j ') $. Then all other members of the family $ [j '] $ have rank $ p $ (see Figure 0). Lemma 2 is proved. 

\begin{figure}
\includegraphics[scale=0.7]{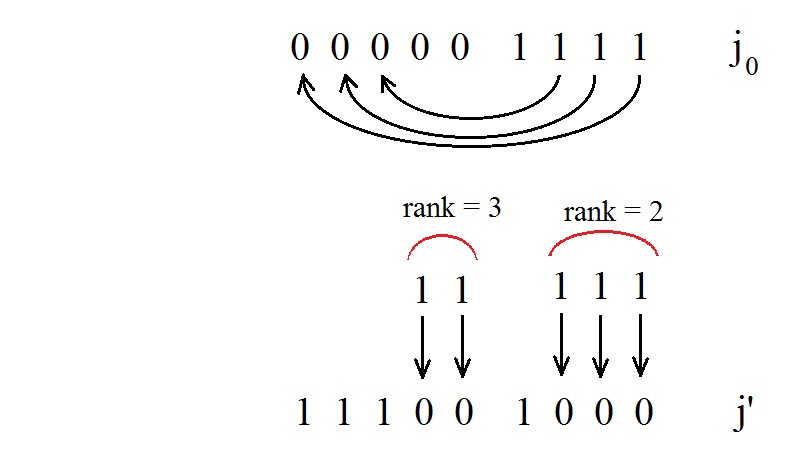}
 \caption{$j'$ - parent of rank 3, obtained from $j_0$ by the substitutions pointed in the upper part of the picture. Two members of its family have ranka 3, nd three members have rank 2: instead of substitution of unit instead of zero in any $q$-th qubit from $ rem\ (j')$ we can omit the substitution with $q$-th qubit in the passage $j_0\rightarrow...\rightarrow j'$ and so obtain instead of $j'$ the new parent of the rank 2 for the member of family $[j']$. }
\end{figure}

We note that, for example, for $ k = n $, there is a unique family, whose parent has rank zero, and this family consists of exactly one member, in which all the qubits have the value one. The rank of this member will also be zero.

We define the amplitude values $\la^0_j$ for all $j\in B^n_k$ depending on the rank $j$ as follows.  Let $p=r(j)$. We put 
\begin{equation}
\la^0_j=(-1)^p\frac{p!}{\prod\limits_{s=0}^{p}(n-k+1-s)}.
\label{la}
\end{equation}

The correctness of this equation follows from Lemma 2, which guarantees the absence of zeroes in the denominator. Indeed, since $p\leq n-k+1$ the only possibility of appearance of such zeroes is the value $s=p=n-k+1$. But the total number of such states $j\in B^n_k$, for which $p=n-k+1$, by Lemma 2 equals zero. 

The equation (\ref{dark_condition}) will not then be true for $j'=j_0$, because the sum of amplitude values for the members of family of rank zero by Lemma 2 is $(n-k+1)/(n-k+1)=1$.  For the members of family of nonzero rank $p$ the equation (\ref{dark_condition}) is satisfied. Really, in view of Lemma 2 in such a family there are exactly $p$ members of rank $p-1$, and exactly $n-k+1-p$ of rank $p$.  
Substituting the amplitude values
$\la^0_j$ from (\ref{la}) for $p$ and $p-1$ we transform the equation \eqref{dark_condition} to the sum of numbers of the form 
$$
(-1)^{p-1}\frac{(p-1)!p}{\prod\limits_{s=0}^{p-1}(n-k+1-s)}+(-1)^p\frac{p!(n-k+1-p)}{\prod\limits_{s=0}^p(n-k+1-s)}=0.
$$

Fulfillment of the equation \eqref{dark_condition} for any family of nonzero rank and its violation for a family of zero rank with the chosen values ​​of variables proves that the equation \eqref{dark_condition} for $ j'= j_0 $ does not depend on other equations of this kind. Since $ j_0 \in B^n_{k-1} $ is arbitrary, all the equations in \eqref{dark_condition} are independent, as required.

The point 1 of the Theorem is proved.

We note that from this point it follows that every state invisible in the RWA approximation is an equilibrium state. Indeed, if the state is dark, then $ 2k \leq n $, because otherwise the dimension of the dark subspace is zero. On the other hand, if the state is transparent, then when zeros are replaced by ones and vice versa, it becomes dark, and we have $ 2k \geq n $, whence $ n = 2k $.

We now prove item 2. Any $ (n, k) $ -singlet can be represented, up to a permutation of qubits, in the following non-normalized form, where the factors of the form $ | 0 \rangle $ are omitted (the number of such factors is $ n-2k $):
\begin{equation}
\begin{array}{ll}
|{\cal S}\rangle=|(1*\ldots**\ldots *0- 0*\ldots**\ldots *&1)(*1\ldots**\ldots 0*- *0\ldots**\ldots 1*)\\
\ldots & (*\ldots*10*\ldots  *- *\ldots*0 1*\ldots *)\rangle
\end{array}
\label{singl}
\end{equation}
which is schematically depicted in Figure 1.

 \begin{figure}
\includegraphics[scale=0.6, bb =-130 -30 160 0]{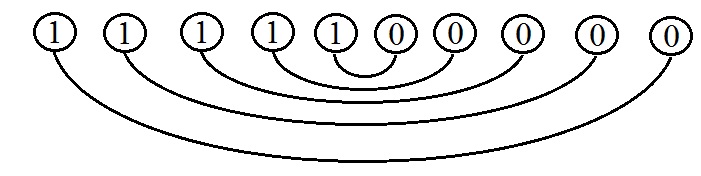}
 \caption{Structure of the singlet state. The tensor product includes all pairs of qubits connected by any arc, so that the values ​​of the qubits are selected either as shown in the figure or in the opposite way. The sign of the pair is positive, if 1 precedes 0 (as indicated in the figure), and negative otherwise}
\end{figure}

The linear span of the set $ A $ is denoted by $ L (A) $, the orthogonal complement to the subspace $ L $ is denoted by $ L^\perp $, the cardinality of an arbitrary set $ A $ is denoted by $ | A | $.

Let $p,q$ be a pair of numbers of qubits, $p\neq q$. Consider the two qubit space $l(p,q)$, generated by the qubits with numbers $ p$ and $ q$, and introduce the following notation for singlet and triplet states in this space:
\begin{equation}
s_{p,q}=|0\rangle_p|1\rangle_q-|1\rangle_p|0\rangle_q,\ t^0_{p,q}=|0\rangle_p|1\rangle_q+|1\rangle_p|0\rangle_q,\  t^1_{p,q}=|0\rangle_p|0\rangle_q,\  t^{-1}_{p,q}=|1\rangle_p|1\rangle_q.
\end{equation}
The first is a singlet, the other three are triplet states. These states form an orthogonal basis in $ l (p,q) $.

Consider an arbitrary state $ | \Psi \rangle \in L (B^n_k) $ and let $ (p,q) | \Psi \rangle $ denote the state obtained from $ | \Psi \rangle $ by permuting the qubits $ p$ and $ q$. We introduce the antisymmetrization procedure for the state $ | \Psi \rangle $ - by the equality
$$
An_{p,q}|\Psi\rangle=|\Psi\rangle-(p,q)|\Psi\rangle .
$$
We note that if $|\Psi\rangle$ was dark then $An_{p,q}|\Psi\rangle$ will be dark as well for all $p,q$. 


By $r(p,q)$ we denote the set of basic states $|r\rangle$ of the set of all atoms but two: $p$ and $q$. We denote by $L_{p,q}$ the subspace ${\cal H}^n_k$, consisting of states of the form $s_{p,q}\otimes|R\rangle$, where $|R\rangle\in L(r(p,q))$. These subspaces in general case are not orthogonal for the different pairs $p,q$.\footnote{It is easy to show that dot product of two states from $D^n_{n/2}$, which are tensor products of EPR- singlets is always some degree of two.}
\bigskip

{\it Lemma 3. 

For $p\neq q$ and $|\Psi\rangle\in L_{p,q}$ the following equalities take place: $Im(An_{p,q})=L_{p,q}$, $ Ker(An_{p,q})=L_{p,q}^{\perp}$  $An_{p,q}|\Psi\rangle=2|\Psi\rangle$.}
\bigskip

{\it Proof.} By the definition, antisymmetrization on $p,q$ always gives a state belonging to $L_{p,q}$. We have: $L_{p,q}^{\perp}$ consists of the states of the form  
$$
t^0_{p,q}|\psi^0\rangle+t^1_{p,q}|\psi^1\rangle+t^{-1}_{p,q}|\psi^{-1}\rangle, 
$$
where $|\psi^s\rangle\in L(p,q))$ for $s\in\{ 0,1,-1\}$. The application of antisymmetrization to such states gives zero. Antisymmetrization applied to the states from $L_{p,q}$, gives their doubling. If $|\Phi\rangle\in Ker(An_{p,q})$, then, since, according to what has been proved, the orthogonal component of the state vanishes by antisymmetrization, and the straight component - doubles, we have $|\Phi\rangle\in (L_{p,q}^{\perp})$. Lemma 3 is proved.

We introduce the projector
 ${\cal P}_{p,q}$ on the subspace $L_{p,q}$ in a natural way:
\begin{equation}
\label{projector}
{\cal P}_{p,q}=\frac{1}{2}\sum\limits_{k\in r(p,q)}|s_{p,q}\otimes k\rangle\langle s_{p,q}\otimes k|.
\end{equation}

Lemma 3 can then be written in an equivalent form as the following Corollary:

\bigskip
{\it Corollary.

$An_{p,q}=2{\cal P}_{p,q}$.}
\bigskip

A state $|D\rangle\in D^n_k,\ k>0$ we call singular if it is orthogonal to all $(n,k)$- singlets. 

To prove part 2 of the theorem, it suffices to show that the singular state must be zero. For this we need a number of additional facts concerning the subspace
$D^n_k$ of the dark states. 
\bigskip 

{\it Lemma 4.

For $k>0$

$D^n_k\subset L(\bigcup\limits_{p\neq q}L_{p,q})$.}
\bigskip

{\it Proof.} 

In this Lemma it is necessary to represent any dark state in the form of a sum of states, in each of which a certain two-qubit singlet presents as a tensor factor. The difficulty here is that singlets are not orthogonal, and two such states may overlap. Therefore, in order to prove this Lemma, we need to consider in more detail the trajectories of individual small portions of the amplitude before they are completely calcelled by virtual emission of a photon.

The action of the group $S_n$ on qubits as their transpositions can be naturally extended to the operators on the whole space of quantum states ${\cal H}$, namely: on the basic states of atoms the transposition a$\eta\in S_n$ acts straightforwardly to the atomic component and leaves the field component unchanged and $\eta\sum\limits_{j_p,j}\la_{j_p,j}|j_p\rangle|j\rangle=\sum\limits_{j_p,j}\la_{j_a,j}|j_a\rangle\eta|j\rangle$.

For the Hamiltonian $aH$, acting on the whole space of states ${\cal H}$ we denote by $G_H$ the subgroup $S_n$, consisting of all transpositions $\tau$ of atomic qubits, such that $[H,\tau]=0$. Let $A\subseteq\{ 0,1,...,2^n-1\}$ be subset of basic states of $n$- qubit atomic system. Its linear span  $L(A)$ we call connected with respect to $H$, if for all two states $|i\rangle,\ |j\rangle\in A$ there exists the transposition of qubits $\tau\in G_H$, such that $\tau (i)=j$. In this case for any basic state of photons $|j_p\rangle$ the subspace $|j_p\rangle\otimes L(A)$ we call connected with respect to $H$ as well. The state $|\Psi\rangle\neq 0$ of $n$- qubit system we call connected with respect to $H$, if it belongs to a connected subspace with respect to $H$; in this case the state of the whole cyctem of the field and atoms of the form $|j_p\rangle\otimes|\Psi\rangle$ we call connected with respect to $H$ as well. 
\bigskip

{\it Proposition.

If $|\Psi\rangle=\sum\limits_j\la_j|j\rangle$ is connected with respect to $H$, then any two columns of the matrix $H$ with numbers $J_1=(j_p,j_1),\ J_2=(j_p,j_2)$ and with arbitrary equal field component $|j_p\rangle$, such that $\la_{j_1}$ and $\la_{j_2}$ are nonzero, differ from each other only by permuting the elements.}

\bigskip

Indeed, for such basic states $j_1$ and $j_2$, according to the definition of the $ H $ -connection, there exists $\tau\in G_H$, such that $j_2=\tau(j_1)$. Columns with numbers $J_1,\ J_2$ consist of the amplitudes of the states $H|J_1\rangle$ and $H|J_2\rangle$, respectively. From the commutation condition, we have $\tau H|J_1\rangle=H\tau |J_1\rangle=H|J_2\rangle$, and this just means that the column $ J_2 $ is obtained from the column $ J_1 $ by permuting elements induced by $ \tau $.
The Proposition is proved.

{\it Example.} We consider Tavis-Cummings Hamiltonian $ H_{TC}^{RWA} $ with zero detuning for $ n $ atoms interacting identically with the field. Then $G_H=S_n$ that can be verified straightforwardly: for the random transposition $\tau=(p,q)$ of two atomic qubits and a basic state of the whole system atoms and field $|J\rangle=|j_p\rangle|j\rangle$ the coinsidence of states $\tau H|J\rangle$ and $H\tau |J\rangle$ follows from the equality of forces of interaction between atoms and field. It means that any transposition of atomic qubits commutes with Hamiltonian. Let $\tilde{\cal H}^n_{k_a,k_p}$ be the linear span of such basic states, in which atomic parts have energy $k_a\hbar\omega$ (contain $k_a$ unitsa), and photonic part is a$|k_p\rangle_{ph}$, where $k_a,k_p$ are natural numbers. Then w$\tilde{\cal H}^n_{k_a,k_p}$ will be connected with respect to $H_{TC}^{RWA}$. 

Our goal is to show that if the state $ |\Psi \rangle $ of the whole system of atoms and field is connected with respect to the Hamiltonian $ H $, then the amplitudes of all the basis states in $ | \Psi \rangle $ can be broken up into small portions - amplitude quanta, so that for each quantum its trajectory will be uniquely determined under the action of the Hamiltonian $ H $ on a small time interval, in particular, it will be uniquely determined, with which exactly other quantum of amplitude it will cancel when summing the amplitudes to obtain the subsequent state in unitary evolution $exp(-iHt/\hbar)$.  

Let $|\Psi\rangle=|j_p\rangle\otimes\sum\limits_j\la_j|j\rangle$ be an arbitrary connected with respect to $H$ state of the whole system. In what follows we will use the notations $|i\rangle$, $|j\rangle$ and $|b\rangle$ for designation of basic states of the whole system of atoms and field, if the opposite is not written directly. 

We introduce the important concept of an amplitude quantum as a simple formalization of the transformation of a small portion of the amplitude in evolution on a small time interval when passing between different basis states. Let $ T = \{+1, -1, + i, -i \} $ be a set of 4 elements, called amplitude types: real positive, real negative, and analogous imaginary. The product of types is determined in a natural way: as a product of numbers. A quantum of amplitude of the size $ \epsilon> 0 $ is a train of the form
\begin{equation}
\label{quanta}
\kappa=(\varepsilon,id, |b_{in}\rangle, |b_{fin}\rangle, t_{in},t_{fin})
\end{equation}
where $|b_{in}\rangle,\ |b_{fin}\rangle$ are two different basic states of the system of atoms and photons, $id$ is a unique identification number that distinguishes this quantum among all others,
 $t_{in},t_{fin}\in T$.  Transition of the form $|b_{in}\rangle\rightarrow\ |b_{fin}\rangle$ is called a state transition
, $t_{in}\rightarrow t_{fin}$ - a type transition. Let's choose the identification numbers so that if they coincide, all other attributes of the quantum also coincide, that is, the identification number uniquely determines the quantum of amplitude. There must be an infinite number of quanta with any set of attributes, except for the identification number. Thus, we will identify the amplitude quantum with its identification number, without further specifying this. We introduce the notation:

$$
t_{in}(\kappa)=t_{in},\ t_{fin}(\kappa)=t_{fin},\ s_{in}(\kappa)=b_{in},\  s_{fin}(\kappa)=b_{fin}.
$$

Transitions of states and types of amplitude quanta actually indicate how this state should change over time, and their choice depends on the choice of the Hamiltonian; the quantum size of the amplitude indicates the accuracy of the discrete approximation of the action of the Hamiltonian using amplitude quanta.

The set $ \theta $ of amplitude quanta of the size $\varepsilon$ is called quantization of the amplitude if the following condition is fulfilled:

{\bf Q}. In the set $\theta$ there is no such amplitude quanta $\kappa_1$ and $\kappa_2$, that their state transitions are the same, $t_{in}(\kappa_1)=t_{in}(\kappa_2)$ and wherein
 $t_{fin}(\kappa_1)=-t_{fin}(\kappa_2)$, and also there are no such quanta of amplitude $\kappa_1$ and $\kappa_2$, that $s_{in}(\kappa_1)=s_{in}(\kappa_2)$ and
 $t_{in}(\kappa_1)=-t_{in}(\kappa_2)$ . 

\bigskip

The condition {\bf Q} means that in the transition described by the symbol "$ \rightarrow $" the final value of the amplitude quantum can not be cancelled with the final value of a similar amplitude quantum.

We introduce the notation $\theta (j)=\{ \kappa:\ s_{in}(\kappa)=j\}$. If $|j\rangle,\ |i\rangle$ are basic states, $t_i,t_j\in T$ are types, $\theta$ is quantization of the amplitude, we introduce the notation
$K_\theta (i,j,t_i,t_j)=\{ \kappa\in\theta(j),t_{in}(\kappa)=t_j,t_{fin}(\kappa)=t_{i}, s_{fin}(\kappa)=i\}$.

For any complex $ z $, we define its relation to the type $ t \in T $ in the natural way: $[z]_t=|Re(z)|$, if $t=+1$ and $Re(z)>0$, or  $t=-1$ and $Re(z)<0$;  $[z]_t=|Im(z)|$, if $t=+i$ and $Im(z)>0$, or $t=-i$ and $Im(z)<0$; $[z]_t=0$ in all other cases.

We call $\theta$- shift of the state $|\Psi\rangle$ the state $|\theta \Psi\rangle=\sum\limits_i\mu_i|i\rangle$, where for every basic $|i\rangle$
 \begin{equation}
\label{shift}
\mu_i=\langle i|\theta\Psi\rangle=\varepsilon\sum\limits_{\kappa\in\theta:\ s_{fin}(\kappa)=i}t_{fin}(\kappa).
\end{equation}
Quantization of amplitude $\theta$ actually specifies the transition $|\Psi\rangle\rightarrow |\theta \Psi\rangle$.

We fix the dimension $ dim ({\cal H}) $ of the state space, and we will make estimates (from above) of the positive quantities: the time and size of the quantum of amplitude to within an order of magnitude, assuming all the constants to depend only on independent constants: $ dim ({ \cal H}) $ and on the minimum and maximum absolute values ​​of the elements of the Hamiltonian $ H $. In this case, the term strict order will mean an estimate from above as well as from below by positive numbers that depend only on independent constants.

We show that for the state $ | \Psi \rangle $ connected with respect to $H$ and for any however small $ \varepsilon> 0 $ there exists $ \delta> 0 $ of strict order $ \varepsilon $ and quantization of the amplitude $ \theta $ with the size of strict order $ \varepsilon^2 $ such that $ \theta $ approximates the state $ | \Psi \rangle $ with error $ \varepsilon $ and the state of the form $ \delta H | \Psi \rangle $ with the same error is approximated by $ \theta $ -shift. Then, passing to the Tavis-Cummings Hamiltonian, we fix the error of our approximation to zero: $ \varepsilon \rightarrow 0 $, so that the overwhelming (for $ \varepsilon \rightarrow 0 $) number of amplitude quanta is cancelled with each other, giving in the limit the state from $L(\bigcup\limits_{p\neq q}L_{p,q})$.

\bigskip

{\it Lemma 4.1. 

Let $|\Psi\rangle$ be a state of the whole system of atoms and field  connected with respect to $H$. 
Then for any number $\varepsilon>0$ there exists the amplitude quantization $\theta$ of the size $\epsilon$ of the order $\varepsilon^2$, the number $\varepsilon_1$ of the order $\varepsilon$ and the number $c$ of the stricked $1$, such that the following conditions are satisfied:

1) for any basic state $j$ 
\begin{equation}
|\epsilon (\sum\limits_{\kappa\in R_+}1-\sum\limits_{\kappa\in R_-}1+i(\sum\limits_{\kappa\in I_+}1-\sum\limits_{\kappa\in I_-}1))-\langle j|\Psi\rangle|\leq\varepsilon
\label{amplitude_expansion}
\end{equation}
where $R_+=\{ \kappa:\ \kappa\in\theta (j),t_{in}(\kappa)=+1\}$, $R_-=\{ \kappa:\ \kappa\in\theta (j),t_{in}(\kappa)=-1\}$, $I_+=\{ \kappa:\ \kappa\in\theta (j),t_{in}(\kappa)=+i\}$, $I_-=\{\kappa:\ \kappa\in\theta (j),t_{in}(\kappa)=-i\}$ and

2) for any basic states $|j\rangle,\ |i\rangle$ and any types $t_{j}, t_{i}\in T$ the following inequality takes place 
\begin{equation}
\label{passage_expansion}
|\epsilon\left(\sum\limits_{\kappa\in K_\theta (i,j,t_i,t_j)}1\right)-c[\langle j|\Psi\rangle\langle i|H|j\rangle]_{t_j}|\leq \varepsilon_1.
\end{equation}
}

{\it Proof. } The meaning of the point 1) is that the quantization of the amplitude gives a good approximation of the amplitudes of the state $ | \Psi \rangle $; the meaning of the point 2) is that this quantization $ \theta $ in the realization of transitions for all quantums of the size $ \varepsilon $ for each gives an approximation with an error of the order $ \varepsilon $ of the state $ cH | \Psi \rangle $ (see Lemma 4.2 Further).

Let there be given a state connected with respect to $ H $
$|\Psi\rangle=\sum\limits_j\la_j|j\rangle$ and a number $\varepsilon>0$. For $|j\rangle$ with nonzero $\la_j\neq 0$ let
\begin{equation}
\la_j=\langle j|\Psi\rangle\approx sign_{re}( \underbrace{\varepsilon+\varepsilon+\ldots +\varepsilon}_{M_j})+sign_{im}i( \underbrace{\varepsilon+\varepsilon+\ldots +\varepsilon}_{N_j} )
\label{quanta_expansion}
\end{equation}
where $sign_{re}\varepsilon M_j+sign_{im}i\varepsilon N_j\approx \la_j$ is the best approximation of the amplitude $\la_j$ with precision $\varepsilon$; $M_j,\ N_j$ are the natural numbers. 
Thus, the point 1) of the Lemma will be almost fulfilled, only without determining the final states $ | i \rangle $ and finite types $ t_i $, which depend on the Hamiltonian.

We approximate each element of the Hamiltonian in the same way as we approximated the amplitudes of the initial state:
\begin{equation}
\label{hamapp}
\langle i|H|j\rangle\approx \pm(\underbrace{\varepsilon+\varepsilon+...+\varepsilon}_{R_{i,j}})\pm i (\underbrace{\varepsilon+\varepsilon+...+\varepsilon}_{I_{i,j}})
\end{equation}
where $R_{i,j},\ I_{i,j}$ are the natural numbers; real and imaginary parts - with accuracy $\varepsilon$ each, and the signs before the real and imaginary parts are chosen proceeding from the fact that this approximation should be as accurate as possible for the selected $\varepsilon$. 

Amplitudes of the resultant state $ H | \Psi \rangle $ are obtained by multiplying all possible expressions \eqref{quanta_expansion} with all possible expressions
 \eqref{hamapp}:

\begin{equation}
\label{mult}
\la_j\langle i|H|j\rangle\approx (sign_{re}M_{j}\varepsilon+i\ sign_{im}N_{j}\varepsilon )(\pm R_{i,j}\varepsilon\pm i\ I_{i,j}\varepsilon).
\end{equation}

Each occurrence of the expression $ \varepsilon^2 $ in the amplitudes of the resultant state after the parentheses are opened on the right side of \eqref {mult} will be obtained by multiplying a certain occurrence of $ \varepsilon $ in the right part of \eqref {quanta_expansion} by a certain occurrence of $ \varepsilon $ in the right part of \eqref {hamapp}. The problem is that the same occurrence of $ \varepsilon $ in \eqref {quanta_expansion} corresponds not to one but several occurrences of $ \varepsilon^2 $ to the result, and therefore we can not associate the amplitude quanta directly with occurrences of $ \varepsilon $ in \eqref {quanta_expansion}. 

How many occurrences of $ \varepsilon^2 $ in the amplitudes of the state $ H | \Psi \rangle $ correspond to one occurrence of $ \varepsilon $ in the approximation of the amplitude $ \la_j = \langle j | \Psi \rangle $ of the state $ | \Psi \rangle $ ? This number, the multiplicity of the given occurrence of $ \varepsilon $, is equal to $ \sum\limits_i (R_{i, j} + I_{i, j}). $ These numbers can be different for an arbitrary Hamiltonian $ H $ and states $ | \Psi \rangle $. However, since $ | \Psi \rangle $ is connected with respect to $ H $, by virtue of the Proposition, the columns of the matrix with different numbers $ j $ for nonzero $ \la_j $ will differ only by permuting the elements, therefore the numbers $ \sum \limits_i (R_{i , j} + I_{i, j}) $ for different $ j $ will be the same.

We introduce the notation $\nu=\sum\limits_i(R_{i,j}+I_{i,j})$ - this is the number of occurences of $\varepsilon$ in any column of the the expansion of the matrix \eqref{hamapp}. The definition of connectivity involves that for any $j=0,1,2,...,N-1$, such that $\la_j\neq 0$ one of numbers $\langle i|H|j\rangle,\ i =0,1,2,...,N-1$ is nonzero, hence for the sufficiently small $\varepsilon$ the number $\nu$ will be nonzero as well and for the sufficiently small $\epsilon$ this number will be of the order $1/\varepsilon$.
\bigskip

We denote by $Z_{i,j}$ the set of occurences of the letter $\varepsilon$ in the right side of the expression \eqref{hamapp}, и пусть $Z_j=\bigcup_iZ_{i,j}$. Then the number of elements in the set $Z_j$ is $\nu$. 

We take the lesser value of amplitude quantum: $\epsilon=\varepsilon/\nu$. We substitute in expression\eqref{quanta_expansion} instead of each occurrence of $\varepsilon$ its formal expansion of the form  $\varepsilon=\overbrace{\epsilon+\epsilon+\ldots +\epsilon}^\nu$, having obtained a decomposition of the amplitudes of the initial state into smaller numbers:

\begin{equation}
\label{refined_expansion}
\begin{array}{ll}
\la_j=&\langle j|\Psi\rangle\approx sign_{re}( \underbrace{\overbrace{\epsilon+\epsilon+\ldots +\epsilon}^\nu+\overbrace{\epsilon+\epsilon+\ldots +\epsilon}^\nu+\ldots +\overbrace{\epsilon+\epsilon+\ldots +\epsilon}^\nu}_{M_j})+\\
&sign_{im}i( \underbrace{\overbrace{\epsilon+\epsilon+\ldots +\epsilon}^\nu+\overbrace{\epsilon+\epsilon+\ldots +\epsilon}^\nu+\ldots +\overbrace{\epsilon+\epsilon+\ldots +\epsilon}^\nu}_{N_j} )
\end{array}
\end{equation}

Let $W^j_1,W^j_2,...,W^j_{M_j+N_j}$ be the sets of occurrences of the letter $\epsilon$ into the right side of the expression \eqref{refined_expansion}, marked with upper braces. Each of these sets has $\nu$ elements, as in the defined above sets $Z_j$. Hence we can build for each such set $W^j_s$ one-to-one mapping of the form $\xi:\ W_s^j\rightarrow Z_j$. For each occurrence of $\varepsilon$ in \eqref{quanta_expansion} we natirally define its descendants - the occurrences of $\epsilon$ in \eqref{refined_expansion}; descendants for each occurrence will be $\nu$.

To each pair of the form $(w_s^j,\xi(w_s^j))$, where $w_s^j\in W_s^j$, we put in correspondence the state and the type transition naturally. Namely, the state transition will be $j\rightarrow i$ for such $i$, that $\xi(w_s^j)\in Z_{i,j}$; the type transition $t_{in}\rightarrow t_{fin}$ is defined so that $t_{in}$ is the type of the occurrence\footnote{The type of an occurrence is also defined naturally, after opening parentheses, for example, for the occurrence $...-i\epsilon ...$ its type is $-i$.}  $w_s^j$, and the type $t_{fin}$ is the multiplication of the type $t_{in}$ by the type of occurrence $\xi(w_s^j)$. The sets $W^j_s$ do not intersect for the different pairs $j,s$, therefore we consider the domain of definition of the function $\xi$ all occurrences of $\epsilon$ in the right side of \eqref{refined_expansion}  (see Figure 2). 

We associate each occurrence of $ \epsilon $ in the expression \eqref{refined_expansion} with a unique identifier and determine its amplitude quantum so that: a) the initial state and initial type of this quantum correspond to this occurrence; and b) the transition and types for a given quantum correspond to the mapping $ \xi $ in the sense defined above. The condition {\bf Q} is satisfied, since there are no cancelling terms in the expression for the matrix element \eqref{hamapp}. Therefore, we determined the quantization of the amplitude.

Then the point 1 of Lemma 4.1 will be fulfilled by the initial choice of the partition \eqref {quanta_expansion}. In view of our definition of the function $ \xi $, the amplitude distribution in the $ | \theta \Psi \rangle $ state will be proportional to the amplitude distribution in the state $ cH | \Psi \rangle $ for any constant $ c> 0 $. In fact, we are talking about the choice of the time value $ t = c $ in the action of the operator $ tH $ on the initial state. In order to determine the value of $ c $ necessary for the fulfillment of the point 2, we calculate the contribution of each occurrence of l $\varepsilon$ in the right side of equation a\eqref{mult} and compare it with the deposit of the corresponding letter $\epsilon$ in $ |\theta\Psi\rangle$. 

We fix some type transition $t_{in}\rightarrow t_{fin}$ and some state transition $s_{in}\rightarrow s_{fin}$. We call an occurrence of $\varepsilon^2$ in the result of opening parentheses in \eqref{mult} corresponding to these transitions if $j=s_{in},\ i=s_{fin}$, and this occurrence is obtained by the multiplication of the occurrence of $\varepsilon$ of the type $t_{in}$ in the first multiplier of the right side of  \eqref{mult} by the occurrence of $\varepsilon$ in the second multiplier of the type $t'$, so that $t_{in}t'=t_{fin}$. Each of such occurrence of $\varepsilon^2$ corresponds to unique quantum of amplitude of the size $\epsilon$ from the amplitude quantization defined above through the function $\xi$, which has the same state anf type transitions: this quantum corresponds to the occurrence of $\epsilon$ that are mapped by the one-to-one correspondence $\xi$ into the initial occurrence of $\epsilon^2$. Hence the target value of $c$ we can find from the proportion $\varepsilon^2/1=\epsilon/c$, whence, taking $\epsilon=\varepsilon/\nu$, we obtain $c=1/\nu\varepsilon$, that has the order 1. 

Since the accuracy of the approximation of the final state by $\theta$- shift coincides in order of magnitude with  $\varepsilon$, we obtain the inequality \eqref{passage_expansion}. Lemma 4.1 is proved. 

\begin{figure}
\begin{center}
\includegraphics[height=0.5\textwidth]{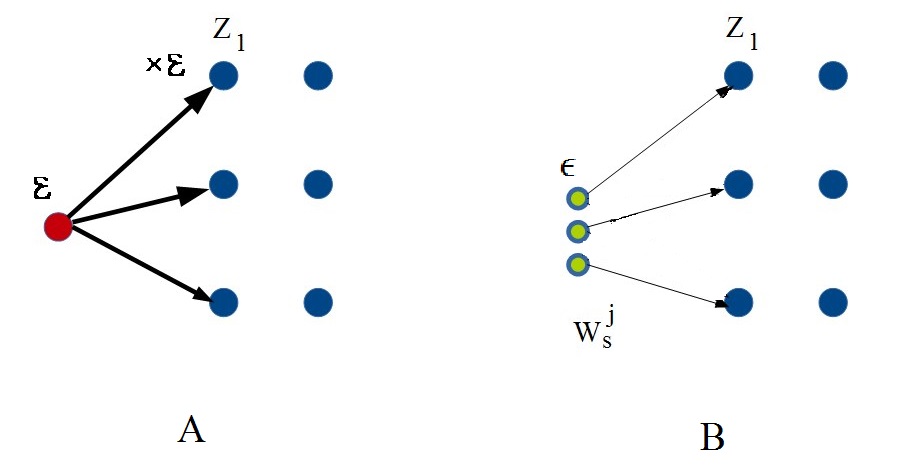}
\caption{A. Multiplication of the state vector by the matrix $H$. The deposit of each occurrence of $\varepsilon$ is multiplied by $\varepsilon$.
B. $\theta$- shift of the initial state. The size of amplitude quantum $\epsilon$ has the order $\varepsilon^2$. }
\label{fig:nps}
\end{center}
\end{figure}

Lemma 4.1 straightforwardly gives
\bigskip

{\it Corollary

In the conditions of Lemma 4.1.  $\|\ |\theta\Psi\rangle-c H|\Psi\rangle\|$ has the order $\varepsilon$.  }
\bigskip

The corollary means that we can assign to each quantum of the amplitude its own history, that is, to assign to it the portion of the amplitude in the state $ c H | \Psi \rangle $, which is in the natural sense the descendant of a given quantum. In particular, we can say that two quanta of amplitude cancel each other when $ \theta $ shift, if their descendants cancel each other.
\bigskip

Now we can prove Lemma 4.

Choose a number $k\in\{ 1,2,...,N-1\}$ and $|D\rangle\in D^n_k$. We consider the subspace $\tilde{\cal H}^n_{k,0}$, defined above. The state $|0\rangle_p |D\rangle\in\tilde{\cal H}^n_{k,0}$ will be connected with respect to Hamiltonian $H=H_{TC}^{RWA}-k\hbar\omega I$, because  $D_k^n\subset {\cal H}^n_k$, all states from $B^n_k$ are obtained from each other by permutations of atomic qubits and all such permutations commute with Hamiltonian $H$ (see the example to the Proposition above). 

Then $H$ coincides with the operator $a^+\bar\s+a\bar\s^+$ on the subspace ${\cal H}=|0\rangle_p\otimes{\cal H}^n_k$,  
e.g. the dark states from $D^n_k$ are the atomic parts of the states from the kernel of $H$, limited on ${\cal H}$. 
Since all atoms interact in the same way with light, we can assume that all nonzero elements of $ H $ are the same, and changing the time scale - that they are equal to one.

 We apply Lemma 4 to the Hamiltonian $H$ and the initial state $|\Psi\rangle=|0\rangle_{p}|D\rangle\in Ker (\bar \s)|_{\cal H}$. For the arbitrary  $\varepsilon>0$ we obtain the approximation of the state $cH|\Psi\rangle$ with the accuracy of the order  $\varepsilon$ by $\theta$- shift for that amplitude quantization $\theta$ with the quantum of the size  $\epsilon$ of the order $\varepsilon^2$ whose existence is asserted in Lemma 4.1.
We have $cH|\Psi\rangle=0$. Further in the transition $|0\rangle_{p}|j\rangle\rightarrow |1\rangle_{p}|i\rangle$ we omit the photonic part.

The Corollary from Lemma 4.1 means that we can expand the amplitudes $ \lambda_j = \langle j | \Psi \rangle $ of the initial state into the sum of the terms $ \pm (i) \epsilon $ so that each occurrence of such a term in the expansion of the amplitude of any basic state $ |j \rangle $ in the state $ | \Psi \rangle $ there will correspond exactly one term of the form $ \pm (i) \epsilon $ in the expansion of the amplitude of some basis state $ | i \rangle $ to the resulting state $ | \theta \Psi \rangle $, this correspondence will be one-to-one, and the transition $ | j \rangle \rightarrow | i \rangle $ will be the emission of a photon, that is, the atomic part state $ | i \rangle $ will be obtained from the atomic part $ | j \rangle $ by replacing one unit with zero.

We combine some occurrences of $ \epsilon $ in the amplitudes of the decomposition of the resultant state into mutually cancelling pairs: $ \pm (i) \epsilon $ corresponding to one basic state. Then the corresponding terms of the initial state will be EPR singlets, since the pair of initial basic states $ | j \rangle $ belongs to the same family, because of the {\bf Q} property of quantization of amplitudes, they are different, and their amplitudes are opposite. Since the difference between $ | \theta \Psi \rangle $ and $ cH | \Psi \rangle = 0 $ ($ c $, of course, depends on $ \varepsilon $) converges to zero for $ \varepsilon \rightarrow 0 $ by \eqref {passage_expansion}, the fraction of the cancelling quanta can be made arbitrarily close to unity as $ \varepsilon $ decreases.

The sum of such pairs of states will belong to a set of the form $ L_{p,q} $, since such a cancellation means the presence of one singlet in the expansion of the basis states. Since there is a fixed number of basic states, letting $ \varepsilon \rightarrow 0 $, we get a sequence of linear combinations of states from $ L_{p,q} $ that converges to some such combination, which is the desired representation of $ | D \rangle $. Lemma 4 is proved.

\bigskip

Let $|D_0\rangle$ be a singular state.
By Lemma 4, we have
\begin{equation}
\label{expansion}
|D_0\rangle= \sum\limits_{p\neq q}s_{p,q}\otimes |D_{p,q}\rangle
\end{equation}
where $|D_{p,q}\rangle$ are the states of $n-2$ qubits. 

Each summand of this sum belongs to the subspace $ L_{p,q} $. The difficulty is that we can not say that $ | D_{p,q} \rangle $ are dark states, that is, the emission of a photon by atoms in any of these states can be compensated by the emission of a photon by an atom whose state belongs to another $ | D_{p ', q'} \rangle $, where $ p '\neq p $ or $ q' \neq q $.

We will overcome this difficulty with the help of an antisymmetrization operation. We put $ |D'_{p,q} \rangle = An_{p,q} | D_0 \rangle $. Then $ | D'_{p,q} \rangle $ for any $ p \neq q $ will be singular, since the darkness and orthogonality of the singlet is preserved under permutation of atoms and subtraction.

We show that there is nonzero among all possible states $ | D'_{p,q} \rangle $. Indeed, let all such states be zero. Then, by Lemma 3, for any pair $ p \neq q$ $ | D_0 \rangle \in L_{p,q}^\perp $, and, the state $ | D_0 \rangle $ belongs to the orthogonal complement of the linear span of all 
$ L_{p,q} $. But in this case it is zero, since it belongs to this linear span by virtue of \eqref{expansion}. 

Thus, among $ | D'_{p,q} \rangle $ there is a nonzero; let it correspond to the pair $ p = 1, q = 2 $: $ | D'_{1,2} \rangle $. This state is singular, and it belongs to $ L_{1,2} $, that is, it has the form $ s_{1,2} \otimes | D_1 \rangle $. Then $ | D_1 \rangle $ is also a singular state of $ n-2 $ qubits. Indeed, $ | D_1 \rangle $ is a dark one, since it was obtained by splitting one $ s_{1,2} $ singlet from the dark state. 
If it is not singular, then it would have a nonzero projection onto the linear span of $(n-2,k)$ singlets obtained by the removing of the first two qubits from the main space. But then multiplying it by one singlet would also have a non-zero projection already on the linear span of $(n,k)$ singlets, which contradicts the singularity of $|D'_{1,2}\rangle$. 

Thus, $|D_1\rangle$ is a singular state of $n-2$ qubits.
We apply the same arguments to it as to $ | D_0 \rangle $, getting singular $ | D_2 \rangle $ from $ n-4 $ qubits, etc. In the end, we get a singular $ D_k $ singlet, which contradicts the definition of the singularity. The Theorem is proved.

\bigskip

Note that if in the RWA approximation the state is dark, but not invisible, then $ n / 2> k $ and in each component of its singlet decomposition there are zero tensor factors of the form $ | 0 \rangle_j $. For an invisible state there are no such zero components, that is, only singlets are present.

So, we see that the dark states in the exact Tavis-Cummings model coincide with the invisible states for this model in the RWA approximation. Indeed, the latter, as follows from the Theorem, are linear combinations of the tensor products of the EPR singlet $ | 01 \rangle- | 10 \rangle $, and each such singlet itself will be dark in the exact Tavis-Cummings model, as is easily seen directly, applying the Hamiltonian $ H_{TC} $ to such an EPR pair. This explains the advantage of the term "dark states": it covers not only those that do not emit light, but also do not absorb light.

The algebraic definition of a dark state for two-level atoms is as follows: $ J_\pm | \Psi \rangle = 0 $, where $ J_\pm $ is an increasing and decreasing operator. It is proved in the paper \cite{K} that this is equivalent to the fulfillment of the inequality $ U^{\otimes n} | \Psi \rangle = | \Psi \rangle $ for any operator $ U \in SU (2) $ (such states $ | \Psi \rangle $ in this work are called ''singlet''). Applying our Theorem, we find that the stationary points of the group $ U^{\otimes n}, \ U\in SU(2) $ are exactly linear combinations of tensor EPR-singlet products, which means the equivalence of the definition of darkness in \cite{K} and our definition of darkness for an exact model.

The work \cite{K} contains a similar algebraic characteristic of the dark states of $ d $ - level atoms is also given for $ d> 2 $; an explicit description of such states is an interesting problem.

\section{Almost dark states}

Consider the state $ | aD \rangle = | 11 \rangle- | 00 \rangle $ of two identical two-level atoms that is not dark, but represents an example of an almost dark state. At low frequencies $ \omega $, this state will persist for a long time, not emitting a photon. Indeed, in the exact Tavis-Cummings model, the transition to the ground state with the emission of a photon for this state can occur in two ways: either the photon is emitted by an excited atom or it arises together with the excitation of another atom in the ground state. It is not difficult to see that the amplitudes of these processes are opposite.

This, however, does not mean that the emission of a photon is impossible at all. The matter is that the excited state $ | 1 \rangle $ and the basic $ | 0 \rangle $ evolve differently: the phase of the excited state changes faster than the ground state, since $ \omega_a> 0 $. Therefore, the states resulting from the emission or production of a photon will differ slightly in phase and there will be no complete cancellation of the amplitudes. This almost dark state differs from the singlet state: in the latter, both transitions are completely equal in both RWA and in the exact model. But if $ \omega_a $ is very small compared to $ g / \hbar $ (the limit of strong interaction, opposite to RWA), then an almost dark state will be at rest for a long time and will not emit a photon.

The tensor product of simple EPR singlets and states of the form $ | aD \rangle $, and linear combinations of such states will also remain unchanged long for small $ \omega $. Is it true that such linear combinations exhaust all states that have the property of almost darkness, that is, of arbitrarily long conservation for small $ \omega $? This question is still open.

\section{Some generalizations}

First, assuming, as before, the equality of forces of interaction with the field of all atoms, we give up the RWA approximation, and consider the case of the exact solution. The set of dark states for the exact Hamiltonian is $Ker (\bar\s+\bar\s^+)=Ker (\bar\s)\cap Ker (\bar\s^+)$, since $\bar \s$ lowers the Hamming weight of the basic states, and $\bar\s^+$ increases it. Given that the replacement of the zeros to ones and vice versa subspaces of $Ker(\bar\s)$ and $Ker(\bar\s^+)$ are moving one to another, and singlet only changes the sign, and applying to $Ker(\bar\s)$ and $Ker(\bar\s^+)$ item 2 of the Theorem, we get that the dark state for the exact Hamiltonian are linear combination of $(2k,k)$ - singlets.  These states will be also invisible. In particular, dark states will exist only for ensembles with an even number of atoms.

Now, on the contrary, we assume that the RWA approximation is true, but the forces of interaction of atoms with the field $g_q$ are different positive real numbers. Now dark subspace is $Ker (\sum\limits_qg_q\s_q)$. Let $s\in B^n_k$ be a binary train, in which zeroes stand on the positions  $s_1,s_2,...,s_k$. We introduce the notations $r_s=\prod\limits_{q\in \{ s_1,s_2,...,s_k\}} g_q$. 
It follows from the definition of Hamiltonian and numbers $r_s$ that the atomic state $|\Psi\rangle=\sum\limits_j\la_j|j\rangle$ is dark if and only if the following system of equations: 
\begin{equation}
\sum\limits_{s\in [j']}r_s\la_s=0, 
\label{dark_condition_def}
\end{equation}
 is satisfied for all $ j'=0,1,\ldots, 2^n-1$, which is connected with the system  \eqref{dark_condition} naturally: $\la^0_s$ is a solution of \eqref{dark_condition_def} if and only if $\la'_s=\la^0_s/r_s$ is a solution of \eqref{dark_condition_def}.

The point 1 of the Theorem is then satisfied because the dimension of the dark subspace does not depend on $g_q$, the point 2 will be also true if only instead of singlet we always consider the ''distributed singlet'': two qubit state of the form $|\bar s_{12}\rangle=g_1|0_11_2\rangle-g_2|1_10_2\rangle$. Such a state is obtained from the singlet by adiabatic change of coordinates of atoms inside the cavity (for example, by optical tweezers), so that the coefficient $g_q$ depends on the coordinate of $q$- th atom (see the first paragraph). 

In this case dark states will not be transparent already when $n=2$, because transparent will be anti-singlet of the form $|(\bar s)^{-1}_{12}\rangle=g_2|0_11_2\rangle-g_1|1_10_2\rangle$. The transparency does not thus connected with the stability of the state in the time in contrast with the darkness, which guarantees such a stability. By the same reason in the case of exact Hamiltonian and the different forces of interaction there is no dark states even for $n=2$. 

\section{Conclusion}

An explicit form of the dark states of an ensemble with an even number of identical two-level atoms in the framework of the Tavis-Cummings model was studied. At the same force of interaction of atoms with light atomic ensembles in these states do not interact at all with the mode of the cavity, and therefore - theoretically - remain unchanged even when the ensemble of atoms is extracted from the resonator. Spatial separation of the dark ensemble or thermal dephasing immediately leads to the emission of photons. Dark states can be used to protect quantum computing, as energy storage, and so on.

The dimension of the dark subspaces is equal to the Catalan numbers. An explicit form of their structure is established: dark states are linear combinations of tensor products of EPR singlet states. Subject to the applicability of the RWA approximation, the dark property is maintained at the vacuum state of the cavity field in the case of adiabatic dilution of atoms, in which the force of interaction with light becomes different. However, such ensembles will interact with light if the state of the field in the cavity is not vacuum.

The search for further applications of dark states and methods for obtaining them is a task for further research. Almost dark states, which are a linear combination of triplets, were also considered; they interact very weakly with light at small values of the excitation energy of atoms, which can be realized, for example, for Rydberg States. Classification of almost dark states as well as dark states in systems of $d$-level atoms at $d>2$ represent separate problems.

In proving the key result of the paper - point 2 of the Theorem, the method of amplitude quanta was developed - small portions of the amplitude of basis states, the trajectory of which can be determined in advance in the course of evolution. This method assumes the passage to the limit, but allows us to prove the algebraic property of dark states. It can be of interest for studying the physics of quantum computers and their scalability.

\section{Acknowledgements}

The work is supported by the Russian Foundation for Basic Research, grant a-18-01-00695.

\end{document}